\documentclass[12pt,a4paper]{article}

\usepackage[margin=2.5cm]{geometry}
\usepackage{times}
\usepackage{graphicx}
\usepackage{amsmath,amssymb}
\usepackage{authblk}
\usepackage[numbers,sort&compress]{natbib}
\usepackage{hyperref}

\usepackage{bm}
\usepackage{xcolor}
\usepackage{amsmath}
\usepackage{amsfonts}
\usepackage{times}

\usepackage{amsmath,amssymb}
\usepackage{authblk}
\usepackage[numbers,sort&compress]{natbib}
\usepackage{hyperref}
\usepackage[english]{babel}
\usepackage{amssymb}
\usepackage[utf8]{inputenc}%
\usepackage{xcolor}
\hypersetup{
    colorlinks=true,
    urlcolor=black,
    linkcolor=black,
    citecolor=black
}
\title{\bfseries Diffusive dark fluids with Planck-2018 and DESI BAO DR2 Measurements}

\author[1,2]{Shambel Sahlu\footnote{shambel.sahlu@nithecs.ac.za}}
\author[1,2]{Amare Abebe\footnote{amare.abebe@nithecs.ac.za}}
\affil[1]{\small Centre for Space Research, North-West University, Potchefstroom 2520, South Africa}
\affil[2]{\small National Institute for Theoretical and Computational Sciences (NITheCS), South Africa}

\date{}

\begin{document}

\maketitle

\begin{abstract}
In this paper, we constrain the diffusive dark fluid cosmological model, wherein energy is transferred between the two dark components through a diffusion process. We extended the work in  Ref. \cite{sahlu2026observational} by employing  Cosmic Microwave Background (CMB) data from the Planck 2018 measurements in combination with Baryon Acoustic Oscillation (BAO) data from the Dark Energy Spectroscopic Instrument (DESI) DR2 (2024). From the results, we found that the discrepancies in \(H_0\) measurements are \(1.90\sigma\) and \(1.72\sigma\) between the Planck 2018 value \((H_0 = 67.4\pm0.5\ \mathrm{km,s^{-1},Mpc^{-1}})\) and our diffusive model values, \(H_0 = 67.27^{+0.54}_{-0.55}\;,\)  and  \(68.35^{+0.31}_{-0.29} \), respectively. We also observe that the effects of the interaction on cosmic evolution and structure formation; we emphasize this by computing the scale-dependent density contrast and the matter power spectrum, compared with the $\Lambda$CDM model.

\end{abstract}
\section{Introduction}
The standard model of cosmology, the so-called $\Lambda$CDM model, explains a wide range of cosmological phenomena successfully; refer to the references  \cite{riess1998observational, perlmutter1999measurements, bull2016beyond,aghanim2020planckx, louis2025atacama, abdul2025desi}. However, modern cosmology increasingly faces significant challenges, as different cosmological measurements have shown that the universe has been experiencing accelerated expansion due to the exotic fluid known as dark energy \cite{riess1998observational,spergel2003first}, which accounts for $\sim$ 70\% of the universe. Another mysterious component of the universe, referred to as dark matter or missing matter, accounts for  $\sim$ 25\% of the universe, posing a challenge for modern physics.   These major dark components of the universe represent puzzles in physics \cite{weinberg1989cosmological,spergel2015dark}. 
The energy density ratio of dark matter to dark energy ($\rho_{dm}/\rho_{de}$, where $\rho_{dm}$ is the energy density of dark matter and $\rho_{de}$ represents the density of dark energy ) can decline more slowly than $a^{-3}$ (where $a$ is the scale factor), indicating that there is an energy exchange between dark matter and dark energy, as supported by the interacting dark sector models. In such models, the interaction term $Q$ accounts for the exchange of energy between dark matter and dark energy without non-gravitational interactions. 
\\
\\
The interaction model has been proposed to address different cosmological probes \cite{di2021interacting, gao2021relieving, van2025compartmentalization, sahlu2026testing}.  For instance, a broad discussion of the influence of the interaction between early dark energy and dark matter, and its effects on structure growth and the CMB, is presented in the Refs. \cite{pu2015early,murgia2016constraints}. In these references, the parameters of the early dark energy model and the coupling between the dark sectors have been constrained using various observational data. The work presented in \citep{zhai2023consistent} also emphasized the role of CMB probes in interacting dark energy models, utilizing data from the Planck satellite, the Atacama Cosmology Telescope, and WMAP (9-year data). According to their results, the values of $H_0$ are modified due to the transfer of energy from dark matter to dark energy. The growth factor and weak lensing signals for a new class of interacting dark energy models are discussed in Ref. \citep{caldera2009growth}. Furthermore, the growth structure of the universe has been explored in \citep{he2009effects}, where the influence of non-vanishing dark energy perturbations on the evolution of dark matter perturbations has been analyzed in detail. In another study, \citep{liu2022dark} conducted detailed N-body simulations to investigate dark energy models, focusing on the formation of cosmic structures and the properties of dark matter halos. Due to the decay of energy from matter to dark energy, structure formation is slowed down, which leads to inconsistencies with observed structure formation. Consequently, the interacting model can be ruled out based on non-linear structure formation, as the interaction term $Q$ is highly sensitive. Similar discrepancies have been reported by \cite{hashim2018cosmic}, who concluded that the universality of the halo mass function is violated in cosmological models featuring direct interactions between dark energy and cold dark matter. Interacting dark energy models are the most promising models for solving the cosmological tensions. In the work \cite{wang2016dark},  an evolving dark energy field with an appropriate non-gravitational interaction with matter is introduced to solve the coincidence problem, since the ratio of matter to dark energy density is expected to decrease rapidly proportional to the scale factor as the universe expands.   The potential of the interacting dark energy model to solve the Hubble tension $H_0$ in km/s/Mpc is reported by \cite{lucca2021multi,wang2016dark}, which is one of the current challenges in the measurements between the direct and indirect measurements. 
\\
\\
In this manuscript, we focus on an interaction model in which energy is transferred between dark matter and dark energy, or vice versa, through a diffusion mechanism \cite{haba2010energy,calogero2011kinetic,benisty2019unification}, as broadly discussed in the work \cite{sahlu2026observational} using late-time measurements. However, the present work provides deeper insight into the constraining of the diffusive dark fluid model by incorporating early-time measurements of the Cosmic Microwave Background (CMB) from the Planck 2018 data, together with recent late-time measurements of Baryon Acoustic Oscillation (BAO) data from the Dark Energy Spectroscopic Instrument (DESI) DR2 (2024). We employ the Monte Carlo Markov chain (MCMC) simulation using the modified \texttt{CLASS}\footnote{\url{http://class-code.net/}} code \cite{blas2011cosmic} together with \texttt{COBAYA}\footnote{\url{https://cobaya.readthedocs.io/en/latest/}} \cite{torrado2021cobaya} for constraining the model parameters.  We further analyze the scale-dependent behavior and the matter power spectrum to investigate the effects of the diffused dark fluid and to highlight its contribution to structure formation.
\\
\\
The paper is organized as follows: In Section \ref{back}, we review the basic equations for the background Universe, focusing on the diffusive dark-fluid system. The perturbation equations are presented in Section \ref{perurbations}.  The constraining of cosmological parameters from MCMC simulations is done in Section \ref{Results}. In this section, the numerical results of the work are broadly explained.  Finally, in Sec. \ref{disc}, the conclusions are drawn up.

\section{Basic equations}\label{back} 
As presented in Refs. \citep{haba2010energy,calogero2011kinetic,benisty2019unification}, the non-conservative energy-momentum tensor for the dark energy  and dark matter components of the fluid  is given by 
\begin{eqnarray*}
    \nabla_\mu T^{\mu\nu}_{de} = -\nabla_\mu T^{\mu\nu}_{dm} = N^\nu  = \gamma_i u^\nu\;,
\end{eqnarray*}
where the current of the diffusion term for that fluid,  $N^\nu_i = \gamma_i u^\nu$,\; $\gamma_i$ represents the number density, and $u^\mu$: $u^\nu u_\nu =-1$ is the four-velocity of the fluids. The 00-component of the interaction term reads  $N^0 = \gamma_i u^0\;, \mbox{where}\; u^0 = \frac{1}{a}(1,0,0,0) $, $N^0 = \gamma_i/a^3$ for a homogeneous Universe, where the full detail is presented in \citep{haba2016dynamics,sahlu2026observational}.

Projecting the above covariant conservation equations onto a homogeneous and isotropic FLRW background yields the continuity equations commonly employed in phenomenological interacting dark energy (IDE) models.In such models, the background conservation equations are commonly written as
\begin{align}
\dot{\rho}_{\rm dm}+3H\rho_{\rm dm} &= Q_{\rm int}, \\
\dot{\rho}_{\rm de}+3H(1+w_{\rm de})\rho_{\rm de} &= -Q_{\rm int},
\end{align}
thus guaranteeing total dark-sector energy conservation.
With this convention, \(Q_{\rm int}>0\) corresponds to energy transfer from dark
energy to dark matter, while \(Q_{\rm int}<0\) corresponds to energy transfer from
dark matter to dark energy. In the diffusive model considered here, the interaction source for each component is $Q^{\rm i}_{\rm int}=\frac{\gamma_i}{a^3}$, with \(\sum_i\gamma_i=0\), so that \(Q^{\rm dm}_{\rm int}=\gamma_{\rm dm}/a^3\)
and \(Q^{\rm de}_{\rm int}=\gamma_{\rm de}/a^3=-Q^{\rm dm}_{\rm int}\).
Identifying the interaction source as $Q^{\rm i}_{int}=\gamma_i/a^3$, the continuity equation for each fluid becomes:
\cite{haba2016dynamics,maity2019Universe}:
\begin{equation}\label{generlconversation}
\dot{\rho_i}+3\frac{\dot{a}}{a}(1+w_i)\rho_i=\frac{\gamma_i}{a^3}\;, \quad \mbox{where}\; i = b,dm,de.
\end{equation}

Here $\gamma_{\rm b}=0$ for baryons, while $\gamma_{\rm dm}$ and$ \gamma_{\rm de}$ are constant diffusion parameters satisfying: $\gamma_{\rm dm}+\gamma_{\rm de}=0\;,$ which guarantees the conservation of the dark-sector energy. From Eq. \ref{generlconversation}, the conservation equation for each fluid becomes 
\begin{eqnarray} \label{eq}
    \dot{\rho}_{de}+3\frac{\dot{a}}{a}(1+w_{de})\rho_{de}=\frac{\gamma_{de}}{a^3}\;,\quad \dot{\rho}_{dm}+3\frac{\dot{a}}{a}\rho_{dm}=\frac{\gamma_{dm}}{a^3}\;,  \; \mbox{and}\quad \dot{\rho}_{b}+3\frac{\dot{a}}{a}\rho_{b}=0\;.
\end{eqnarray}
We assume the dark energy, where the pressure is taken $p_{de} = w_{de}\rho_{de}$, with equation-of-state parameter $w_{de} = -1$.  A detailed mathematical expression is presented in Ref. \cite{sahlu2026observational}. The above diffusive solution of Eq. \eqref{eq} leads to 
\begin{eqnarray}
H^2 = \frac{8\pi G}{3c^4} \Bigg[ \rho_{\rm m0}a^{-3} + \frac{\gamma_{\rm {dm}}}{H_0}\left(a-1\right)a^{-3} +  \rho_{de0}  -\frac{\gamma_{de}}{2H_0}\left(a^{-2}-1\right)\Bigg]\;,
\end{eqnarray}
where $\rho_{m0} = \rho_{\rm b0}+ \rho_{\rm dm0}$. The dimensionless dynamical quantities, see \cite{sahlu2026observational}:
\begin{eqnarray*}\label{Eq:h_z_1}
  && \Omega_{i}\equiv \frac{8\pi G}{3H^2_0}\rho_{i}\;, Q_{dm}\equiv\frac{8\pi G}{3H^3_0}\gamma_{dm} = - Q_{de}\;,  h\equiv \frac{H(z)}{H_0}\;.  
\end{eqnarray*}
We proceeded with further analysis using the above background evolution equations together with the following perturbation quantities presented in the following section.  


\section{Linear perturbations in diffusive cosmology}\label{perurbations}
In the conformal-Newtonian gauge, the perturbed metric is
\begin{eqnarray}
    ds^2 = a^2(\tau)\Big[ -(1+2\Phi)d\tau^2 + (1-2\Psi)\delta^{ij}dx_idx_i\Big] \;,
\end{eqnarray}
where $\Phi$ and $\Psi$ are the gravitational potentials. In the perturbed universe, the stress-energy-momentum tensor of the  diffusive fluid is given as
\begin{equation}\label{diffusivex}
\delta \left(T^{\mu\nu}_i{}_{;\mu}\right)= \delta N^\nu_i\;, \qquad \nabla_\mu \delta T^{\mu\nu}_{de} = -\nabla_\mu \delta T^{\mu\nu}_{dm} = \delta\left(N_{i}^\mu\right)\;,
\end{equation}
where the diffusion term can be expressed  $N^\mu_i = Q^\mu_{int}(i) u^\mu$, where the interacting diffusive term $ N^\mu_i +\delta N^\mu_i\equiv \delta Q^\mu_{int}(i) u^\mu +Q^\mu_{int}(i)\delta u^\mu$, where $u^0 = \frac{1}{a}\left(1-\Phi\right)$ and $u^i = {\partial_iv^i}$ and the velocity $\theta_i = {\partial_jv^j_i} $. By taking into account  Eq. \eqref{diffusivex}, the perturbed quantities can be expressed: interaction term seems  $\left(\delta Q^0_{int}(i) = -\frac{\gamma_i}{a} \Phi, \;\;  \delta Q^i_{int}(i) = -\frac{\gamma_i}{a}v^i\right)$, and the four velocity vector $u^\mu  =  \tilde{u^\mu} + \delta u^\mu$, the pressure of the fluid  $p_i = \tilde{p}_{i} +\delta p_i$, the density perturbation  $\rho_i = \tilde{\rho}_{i} +\delta \rho_i$,  for each species.
The density perturbation at a comoving location $x$ is most conveniently characterized by its fractional difference $\delta(x,t)$ with respect to the background Universe, $ \delta_i(x,t) \equiv \frac{\rho_i - \tilde{\rho}_{i}}{\tilde{\rho}_{i}}.$

From Eq. \eqref{diffusivex}, the $00$ components of the energy momentum tensor is  $\delta T^{00}_i =  -\delta \rho_i $, the $0-i$ components $\delta T^{0j} = (\rho_{i}+pi)v_{i}^j$ and the $ij$ components are $\delta T^j_i = \delta p_i \delta^j_i$. The generalized form of the density contrast and momentum  conservation equations are \cite{valiviita2008large,majerotto2010adiabatic}

\begin{align}\label{theta}
 {\delta'_{i}}   & +3\mathcal{H}(1+c_{si}^{2})\delta_{i}-(1+w_{i})(3{\Phi'}-\theta
_{i})  +  9\mathcal{H}((1+w_{i})(c^2_{s(i)} - w_i))   =\frac{a}{\rho_{i}}  \delta Q^0_{int}(i)    \nonumber\\&  +\frac{aQ^0_{int}(i)}{\rho_i}\left[ \Psi - \delta_{i} + 3\mathcal{H} (c^2_{s(i)} - w_i)\frac{\theta_i}{\kappa^2} \right]
\;, ~~\rm and \\
&  {\theta'_{i}}+\mathcal{H}(1-3w_{i})\theta_{i}-{k^{2}\Phi%
} - \kappa^2\frac{c^2_{si} \delta_{i}}{(1+w_{i})}=  \frac{ak^{2}}{(1+w_{i})\rho_{i}}\delta Q^j_{int}(i)+\frac{Q^0_{int}(i)}{\rho_i(1+w_i)}(\theta_{i}- (1+c^2_{si})\theta_{i})\;,\label{delta}
\end{align}
respectively. 

In the recent work \cite{sahlu2026observational}, the effect of the density contrast of dark energy is highlighted as being neglected.  This work emphasizes that the dominant role of matter perturbations in having a significant effect on gravitational collapse, which is the cause of structure formation, rather than that of dark energy, \(\delta_{de} \approx 0\). This is because dark energy ($w = -1$) has no perturbations, and for canonical scalar-field dark energy \(c_{si}^2 \approx 1\), pressure support smoothes out perturbations on subhorizon scales. In the current manuscript,  we assume the coupling of the interaction term $N^\mu_i$ is purely timelike, and then \(\delta N^j_i = 0\), so dark energy remains effectively homogeneous.
We also assume that, for this fluid, we only consider the energy transfer and neglect the momentum transfer, where the energy transfer in the dark matter fluid affects the background and density contrasts more directly than the momentum transfer, see more in the  Ref. \cite{valiviita2008large}.   We shall take into account Eqs. \eqref{theta} and \eqref{delta} for further discussion of the structure formation in diffusive dark fluid cosmology in the next section.  
\section{Results and Discussion}\label{Results}
\begin{figure}[h!]
\centering
    \includegraphics[width=1.0\linewidth]{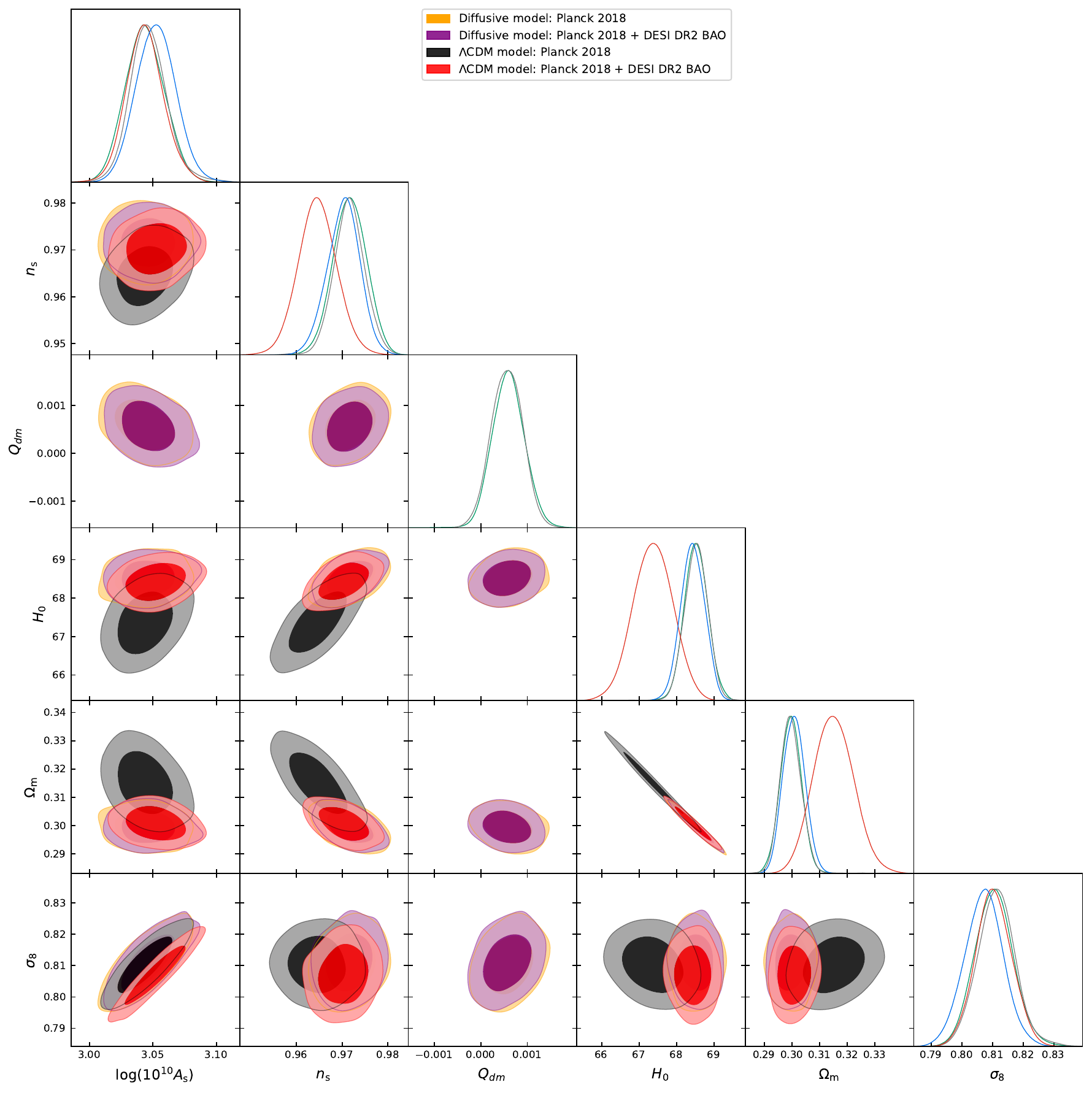}
    \caption{ The MCMC output for diffusive and $\Lambda$CDM models at 68\%  level of  confidence space for the posterior parameters for \texttt{Planck-2018} and  \texttt{Planck-2018+DESI DR2 BAO} measurements}
    \label{fig:placeholder1}
\end{figure}

In this section, we have implemented the modified \texttt{CLASS} code \cite{blas2011cosmic} together with \texttt{COBAYA} \cite{torrado2021cobaya} to constrain the diffusive dark fluid model for further analysis. To constrain the cosmological parameters of the diffusive model, the following measurements are considered: 

\begin{itemize}
\item \textbf{CMB:} From the \textit{Planck} 2018 legacy release, we use the high-$\ell$ \texttt{Plik} TT likelihood ($30 \leq \ell \leq 2508$), TE and EE ($30 \leq \ell \leq 1996$), low-$\ell$ TT ($2 \leq \ell \leq 29$), low-$\ell$ EE ($2 \leq \ell \leq 29$), and the CMB lensing power spectrum \cite{aghanim2020planck}. We refer to this dataset as \texttt{Planck2018}.

\item \textbf{BAO:} We use the DR2 BAO distance and correlation measurements from the Dark Energy Spectroscopic Instrument (DESI) \cite{andrade2025validation, abdul2025desi}. These include isotropic BAO measurements $D_V(z)/r_d$ and anisotropic measurements $D_M(z)/r_d$ and $D_H(z)/r_d$, as well as their correlations. Hereafter, we refer to this dataset as \texttt{DESI DR2 BAO}.
\item \textbf{Joint datasets}: The combined analysis of the data improves the precision and constraints on the models we employ, providing a more complete picture of the Universe; for the joint analysis, we use \texttt{Planck2018}+\texttt{DESI DR2 BAO}, for which the corresponding chi-square is
\begin{eqnarray}
    \chi^2_{\text{sum}} =   \chi^2_{\texttt{Planck2018}} +  \chi^2_{\texttt{DESI DR2 BAO}}\;.
\end{eqnarray}
\end{itemize}
The contour plots of the constrained values of the considered cosmological parameters are presented in Fig. \ref{fig:placeholder1} using \texttt{Planck-2018} and  \texttt{Planck-2018+DESI DR2 BAO} measurements by taking into account the parameter's priors as follows: $H_0$ [km/s/Mpc] $\in[0.5,1.0]$; $\Omega_{m}\in[0.0,1.0]$; $Q_{dm}\in[-0.1,0.1]$; $n_{s}\in[0.8,1.2]$; $\log{A}\in[1.61,3.91]$; $\sigma_{8}\in[0.0,0.1]$; and  $\theta_{s{100}}\in[0.5,10]$. Similarly, Table \ref{tab:table2} summarizes the 95\% level of confidence of parameter values for  $\Lambda$CDM and diffusive dark fluid models.
\\
\\
\begin{table}[h!]
\begin{tabular}{ccc|cc}
\hline
\textbf{Parameters} & \multicolumn{2}{c}{\textbf{Diffusive model}} &  \multicolumn{2}{c}{\textbf{$\Lambda$CDM} model}   \\
\hline
& \texttt{Planck 2018} & \texttt{Planck 2018 + } & \texttt{Planck 2018  } & \texttt{Planck 2018 +
} \\
&&\texttt{DESI DR2 BAO}&&\texttt{DESI DR2 BAO}\\
\hline
\hline
$\log{(A_s)}$ & $3.044^{+0.016}_{-0.016}$ & $3.047^{+0.013}_{-0.015}$& $3.046^{+0.017}_{-0.018}$ & $3.047^{+0.015}_{-0.016}$ \\
$n_s$ & $0.9717^{+0.0037}_{-0.0036}$ & $0.9715^{+0.0034}_{-0.0032}$ & $0.9643^{+0.0047}_{-0.0038}$ & $0.9707^{+0.0042}_{-0.0031}$ \\
$Q_{\rm dm}$ & $0.0005904^{+0.00034}_{-0.00034}$ & $0.0005629^{+0.00033}_{-0.00034}$ & --- & --- \\
$H_0$ & $68.52^{+0.310}_{-0.310}$ & $68.51^{+0.38}_{-0.41}$ & $67.32^{+0.52}_{-0.51}$ & $68.31^{+0.38}_{-0.33}$ \\
$\Omega_m$ &  $0.2996^{+0.0038}_{-0.0038}$ & $0.2996^{+0.0035}_{-0.0041}$& $0.3154^{+0.0075}_{-0.0071}$ & $0.3022^{+0.0038}_{-0.0042}$ \\
$\sigma_8$ & $0.8105^{+0.006}_{-0.0064}$ & $0.8113^{+0.0059}_{-0.006}$ & $0.811^{+0.0079}_{-0.0075}$ & $0.806^{+0.0063}_{-0.0056}$ \\
&&&&\\
\hline\hline

 & \multicolumn{2}{c}{\textbf{Statistical values}} &    \\
\hline
$\chi^2$ & 2775.85 &  2791.88 & 2776.25 &2793.33\\
AIC/$\Delta$AIC &2789.85/1.599 &2805.88/0.080  & 2788.25/--- &2805.33/--- \\
BIC/$\Delta$BIC &2830.62/7.420 & 2846.69/5.980 & 2823.20/--- & 2840.31/--- \\
\hline
\end{tabular}
\caption{ The table shows the parameter values for  $\Lambda$CDM and diffusive dark fluid models at 68\% level of confidence and the statistical values of the models.}
\label{tab:table2}
\end{table}
In the present work, we emphasize how intractable dark energy affects the evolution of the universe through a diffusion mechanism. The parameter $Q_{dm}$ is the key quantity that indicates the flow of energy in the system. As presented in Table \ref{tab:table2}, the values of $Q_{\rm dm}$ are  $0.0005904^{+0.00034}_{-0.00034}$ and $0.0005629^{+0.00033}_{-0.00034}$  for the \texttt{Planck 2018} and \texttt{Planck 2018 + DESI DR2 BAO} datasets, respectively. The sign of $Q_{dm}$ indicates the direction of the energy flow, as presented in \cite{sahlu2026observational}. For instance, when $Q_{dm}<0$, the flow of energy is from dark matter to dark energy, and the opposite conclusion is true. Based on the results presented in Table \ref{tab:table2}, these are the best-fit values of $Q_{dm}>0$, indicating the flow of energy is from dark energy to dark matter for the considered measurements. This result is also consistent with the recent finding \cite{escobal2026linear,FIGUERUELO2026102238}. 

From this table, we note that the deviation between our diffusive model values of \(H_0 = 68.52^{+0.310}_{-0.310}\)  and  \(68.51^{+0.38}_{-0.41}\) and the Planck 2018 result \(H_0 = 67.4\pm 0.5\) km/s/Mpc \cite{aghanim2020planck} is \(1.90\sigma\) and \(1.72\sigma\), respectively. For the \(\Lambda\)CDM model, \(H_0 = 67.32^{+0.52}_{-0.51}, ~~ \&~~ 68.31^{+0.38}_{-0.33}\) corresponds to \(0.120\sigma\) and \(1.519\sigma\). Similarly, the sigma deviations of the \(H_0\) values taken from the Table \ref{tab:table2} and the  SH0ES measurement \(H_0 = 73.04 \pm 1.04\) \cite{riess2019large} are \(4.17\sigma\) and \(4.09\sigma\) for the diffusive model, and \(4.91\sigma\) and \(4.27\sigma\) for the \(\Lambda\)CDM model. Indeed, the deviation between the SNIa (SH0ES) and Planck 2018 measurements of \(H_0\) is \(5.28\sigma\). Using the same models, the sigma deviation is presented \cite{sahlu2026observational} and reported that the diffusive model is favored by Planck 2018 measurements, while the $\Lambda$CDM model is favored by SH0ES measurements.  In the current study, both models are statistically consistent with Planck 2018 measurements but show strong tension with SH0ES, with deviations \(\ge .27\sigma\). A full analysis will be considered in upcoming work with more extensive data sets, including \texttt{Planck 2018 + SNIa} compilations.

In this work we also highlight the statistical analysis of the diffusive model by calculating the minimum Chi-squre (\(\chi^2_{min}\)) Akaike Information Criterion ($\rm{AIC}=\chi^{2}_{min}+2K$) and the Bayesian/Schwarz Information Criterion ($\rm{BIC}=\chi^{2}_{min}+K\log(N_i)$), where $K$ corresponds to the number of free parameters for that specific model, and $N_i$ stands for the number of data points for the $i^{\rm th}$ dataset, see Table \ref{tab:table2}. From Table \ref{tab:table2}, the values of $\Delta$AIC $\leq2$, indicate that the model has substantial support for the observational evidence.\footnote{ $\Delta\rm AIC = \Delta\rm AIC_{diffusive}  - \Delta AIC_{\Lambda CDM} $} For more information see Ref. \cite{Sahlu:2024pxk}. Whereas $\Delta \rm BIC$ value found in the range of $6\leq\Delta\rm{BIC}\leq10$, which indicates the model has weak observational support. However, this statistical penalization comes from the large number of data points and the model's extra degree of freedom, not the physical viability of the considered model. Beyond information criteria, discrimination within the interacting dark-energy class can be sharpened by demanding simultaneous consistency at the background ($H_0$), growth-amplitude ($\sigma_8$) and perturbation-coupling levels. This view, that cosmological tensions act as consistency conditions on a given model, has been formalised as the cosmological consistency triangle \cite{abebe2026consistency} and applied to modified-gravity deviations \cite{abebe2026tensions}.

\begin{figure}[h!]
    \centering
    \includegraphics[width=0.7\linewidth]{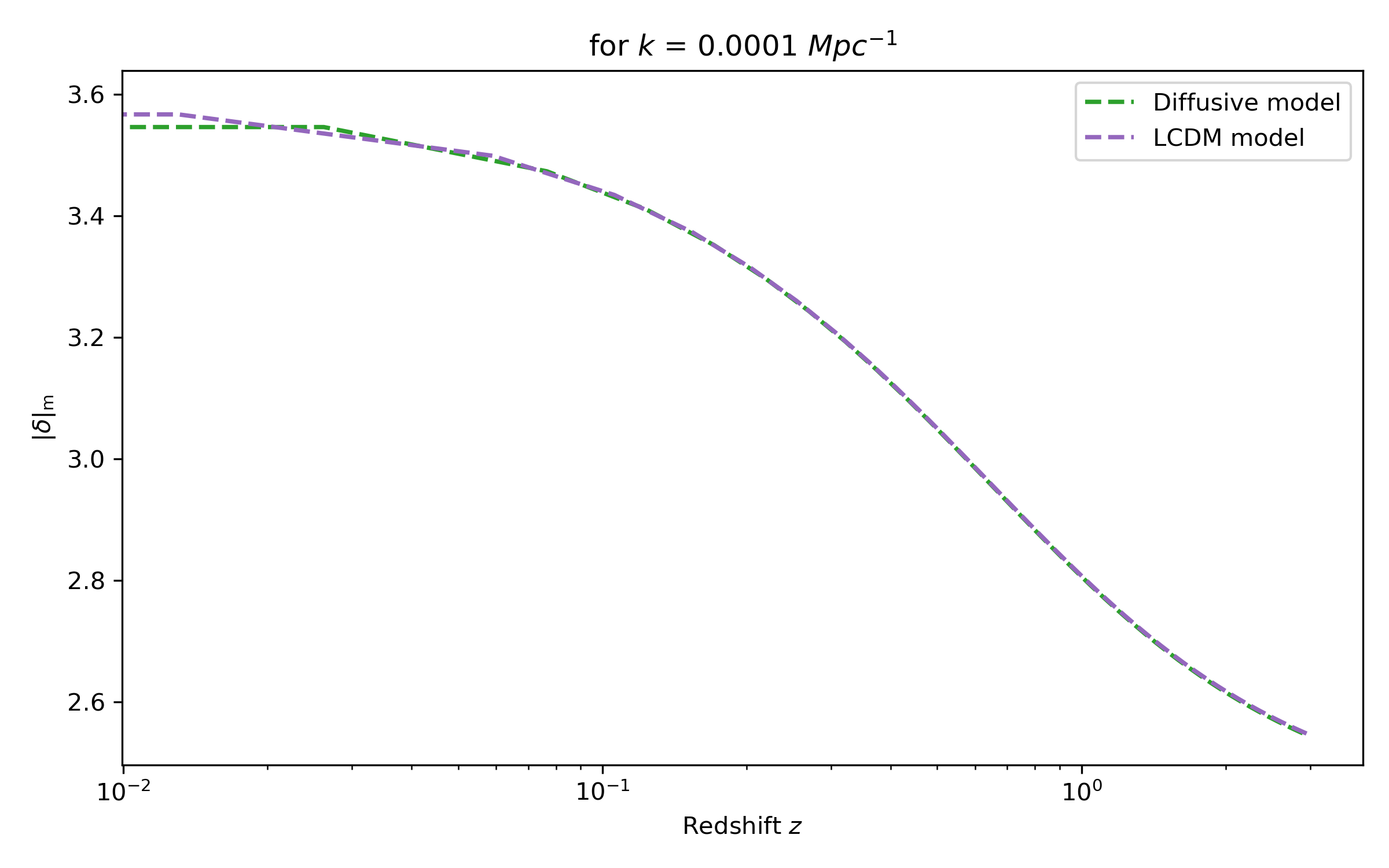}
    \caption{The scale-dependent density contrast \(|\delta|_{m}(z,k)\) for diffusive dark fluid and $\Lambda$CDM models. }
    \label{fig:placeholderdensity}
\end{figure}

By accounting for the constraint values for the cosmological parameters taken from Table \ref{tab:table2}, we present numerical results for the scale-dependent density contrast $|\delta|_m(z,\kappa)$ and the matter power spectrum $P_m(k,z)$ for the diffusive dark-fluid model and compare them with $\Lambda$CDM. First, we show numerical results for the scale-dependent density contrast $|\delta|_m(z,k)$ (see Fig.\ref{fig:placeholderdensity}) for the diffusive dark-fluid model with wave-mode scale $k=0.0001\ \mathrm{Mpc}^{-1}$. The scale-independent density contrast for the same model is discussed in \cite{sahlu2026observational}, where $|\delta|_m(z,k)=|\delta|_m(z)$. In this plot, we clearly see that the diffusive dark-energy model exhibits small deviations from $\Lambda$CDM near $z\approx0$, indicating that energy transfer via diffusion affects late-time cosmic evolution.  
\\
\\
We secondly present the numerical results of the matter power spectrum $P_m(k, \;z = 0,1)$ at redshifts $z = 0$ and $1$ at left panel of  Fig \ref{fig:placeholderpk}. From this plot, the amplitude of the power spectrum is generally higher than that of the diffusive fluid model due to the energy transfer through diffusion interaction.  This deviation of the power spectrum   $\Delta P_m(k, \;z = 0,1)$ is more clearly noticeable in Fig \ref{fig:placeholderpk}, where the deviation is 
$$ \Delta P_m(k, \;z = 0,1) = \frac{P^{\text{diffusive}}_m(k, \;z = 0,1) - P^{\Lambda\text{CDM}}_m(k, \;z = 0,1)}{P^{\Lambda\text{CDM}}_m(k, \;z = 0,1)}\;. $$
From the numerical results of  \(\Delta P_m(k, \;z = 0,1) \) in the right panel of  Fig. \ref{fig:placeholderpk}, we observe that the effect of the interaction for the matter fluctuations across different $k$ ranges. At least the following three cases are noticeable: 

\begin{itemize}
    \item For the case of small scale  $k \gtrsim 10^{-1} h\,\mathrm{Mpc}^{-1}$, we observe that oscillation curves of the matter spectrum, $\Delta P_m(k, \;z = 0,1)$  due to BAO-like wiggles),  where the diffusive model enhance of the matter fluctuations than $\Lambda$CDM  were more clustering in this scale. 
   
\begin{figure}[h!]
    \includegraphics[width=0.48\linewidth]{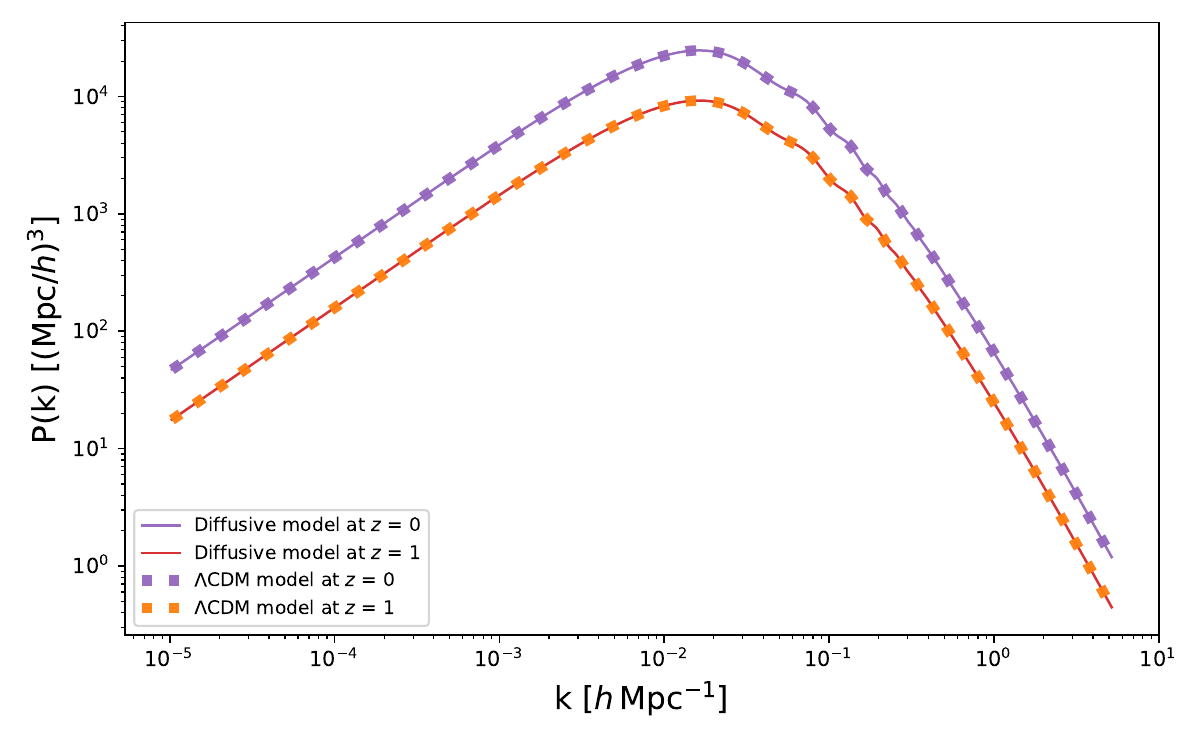}
    \includegraphics[width=0.48\linewidth]{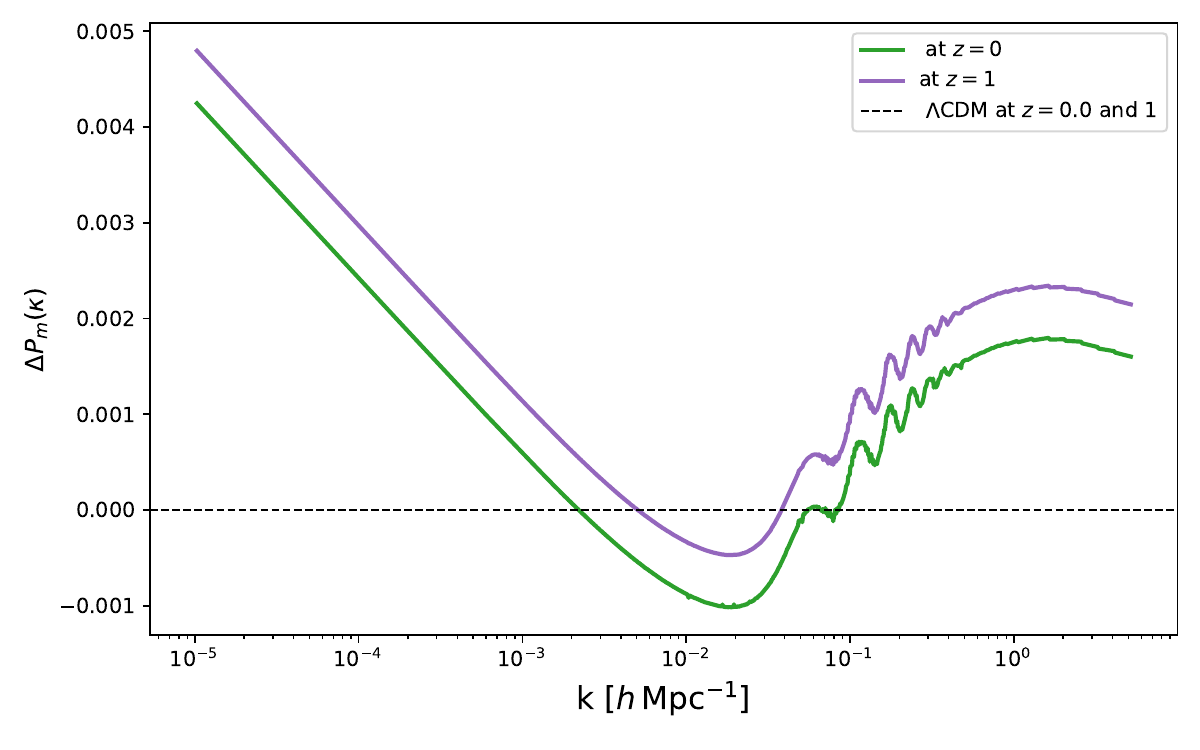}
    \caption{Left panel: the matter  power spectrum $P_m(k, \;z = 0,1) $ for diffusive dark fluid and $\Lambda$CDM models. Right panel: the deviation of matter power spectrum $\Delta P_m(k, \;z = 0,1) $ between diffusive dark fluid and $\Lambda$CDM models.  }
    \label{fig:placeholderpk}
\end{figure}
\item For the case of intermediate scale around $ 10^{-2} \lesssim k \lesssim 10^{-1} h\,\mathrm{Mpc}^{-1}$, we clearly see the monotonically suppressed matter fluctuation against $\Lambda$CDM.  
\item For the case of large scale $k \lesssim 10^{-2} h\,\mathrm{Mpc}^{-1}$, the deviation of the power spectrum   $\Delta P_m(k, \;z = 0,1)$ is nearly smooth and monotonically increased the amplitude of the  $\Delta P_m(k, \;z = 0,1) $, where the diffusive model enhancing the matter fluctuations at this range as well. In general, the amplitude of the $P_m(k) $  at $z = 1$ is higher than at $z =0$ across all scales.  
\end{itemize}

\section{Conclusions}\label{disc}
In this manuscript, we revised the interacting dark energy through a diffusion mechanism with the \texttt{Planck-2018} and \texttt{DESI BAO DR2} measurements. After we highlighted the basic equations of the conservation equation, we demonstrate that the MCMC simulation can constrain the cosmological parameters using the modified \texttt{CLASS} code together with \texttt{COBAYA}.  We have highlighted the $H_0$ tension based on the Hubble parameter values taken from Table \ref{tab:table2} and reported that the diffusive model has a statistical consistency with Planck 2018  but strong tensions with SH0ES.


The best-fit values of \(Q_{\rm dm}\) are positive, whereas Ref.~\cite{sahlu2026observational}
reported negative values using late-time background and growth data. Under the common
sign convention adopted in both analyses, \(Q_{\rm dm}<0\) corresponds to energy transfer
from dark matter to dark energy, while \(Q_{\rm dm}>0\) corresponds to transfer from dark
energy to dark matter. However, the credible intervals obtained here include
\(Q_{\rm dm}=0\), so the apparent difference in the best-fit sign should not be interpreted
as evidence for a statistically established reversal of the energy-transfer direction. It may
instead reflect differences in the datasets, parameter degeneracies, or the treatment of the
background and perturbation sectors. Including  the combinations of early- and late-time measurements makes the results more reliable than work done using only late-time measurements.

The detailed statistical analysis of the model is highlighted by calculating the $\chi^2_{\rm min}$, $\Delta$AIC, $\Delta$BIC to test the good fit of the model with the observations. From the results, the model has been strongly supported according to $\Delta$AIC, but weaker support from the $\Delta$BIC point of view.   The work also emphasized the numerical results of the scale-dependent density contrast and the matter power spectrum. We also present the deviation of the power-spectrum $ \Delta P_m(k, \;z = 0,1)$ that clearly showed the significant effects of the diffusive interaction against $\Lambda$CDM. In the upcoming work, the comprehensive analysis with intensive datasets will be conducted to see the overall picture of the diffusive dark fluid model that possibly alleviates the cosmological tensions.

\end{document}